# Evaluating Ensemble and Deep Learning Models for Static Malware Detection with Dimensionality Reduction Using the EMBER Dataset

Md Min-Ha-Zul Abedin[1][0000-1111-2222-3333] and Tazqia Mehrub[2][1111-2222-3333-4444]

[1] Auburn University, Auburn AL 36849, USA
[2] Bangladesh University of Professionals, Dhaka, Bangladesh
abedinm@auburn.com

**Abstract.** This study investigates the effectiveness of several machine learning algorithms for static malware detection using the EMBER dataset, which contains feature representations of Portable Executable (PE) files. We evaluate eight classification models: LightGBM, XGBoost, CatBoost, Random Forest, Extra Trees, HistGradientBoosting, k-Nearest Neighbors (KNN), and TabNet, under three preprocessing settings: original feature space, Principal Component Analysis (PCA), and Linear Discriminant Analysis (LDA). The models are assessed on accuracy, precision, recall, F1 score, and AUC to examine both predictive performance and robustness. Ensemble methods, especially LightGBM and XGBoost, show the best overall performance across all configurations, with minimal sensitivity to PCA and consistent generalization. LDA improves KNN performance but significantly reduces accuracy for boosting models. TabNet, while promising in theory, underperformed under feature reduction, likely due to architectural sensitivity to input structure. The analysis is supported by detailed exploratory data analysis (EDA), including mutual information ranking, PCA or t-SNE visualizations, and outlier detection using Isolation Forest and Local Outlier Factor (LOF), which confirm the discriminatory capacity of key features in the EMBER dataset. The results suggest that boosting models remain the most reliable choice for high-dimensional static malware detection, and that dimensionality reduction should be applied selectively based on model type. This work provides a benchmark for comparing classification models and preprocessing strategies in malware detection tasks and contributes insights that can guide future system development and real-world deployment.

## 1 Introduction

The escalating sophistication of malware and the increasing velocity at which new variants emerge have rendered traditional signature-based detection mechanisms insufficient. These legacy systems struggle to detect novel and obfuscated threats due to their reliance on pre-defined patterns, making them ineffective against polymorphic and zero-day attacks. To address these limitations, machine learning (ML)-based techniques have gained traction, offering the ability to generalize from known threats and detect previously unseen malware based on learned patterns.

Static malware analysis, which inspects executable files without execution, has emerged as a particularly scalable and safe approach. Among the various static datasets available, the EMBER (Endgame Malware Benchmark for Research) dataset stands out due to its high-quality labeling, rich feature set derived from PE (Portable Executable) files, and accessibility for benchmarking. It enables robust evaluations of diverse ML classifiers in a controlled and reproducible setting.

While numerous studies have demonstrated the effectiveness of machine learning in static malware detection, comprehensive comparisons across model architectures and preprocessing strategies remain limited. In particular, the influence of dimensionality reduction techniques such as Principal Component Analysis (PCA) and Linear Discriminant Analysis (LDA) on classification performance has not been sufficiently explored in the context of malware detection. Moreover, there is a need to assess trade-offs between model complexity, interpretability, and computational efficiency.

This study addresses these gaps by conducting an extensive benchmark of classical, ensemble, and deep learning models using the EMBER dataset under three distinct preprocessing scenarios—no feature reduction, PCA, and LDA. Through rigorous evaluation on multiple metrics and detailed exploratory data analysis, we present an empirical framework for selecting optimal classifiers and feature handling strategies in malware detection pipelines. Contributions:

- Systematic Benchmarking: Evaluated a suite of ML classifiers (LightGBM, XGBoost, CatBoost, Random Forest, Extra Trees, HistGradientBoosting, KNN, and TabNet) on the EMBER dataset using three dimensionality reduction strategies: none, PCA, and LDA.
- Multi-Metric Evaluation: Assessed performance using five classification metrics—accuracy, precision, recall, F1-score, and AUC to provide a comprehensive view of each model's strengths and weaknesses.
- Exploratory Data Analysis: Applied mutual information ranking, outlier detection (Isolation Forest, LOF), and class separability visualization (PCA, t-SNE) to understand feature behavior and its effect on classification.
- Comparative Model Insights: Found that tree-based ensemble models, particularly LightGBM and XGBoost, outperformed deep learning (TabNet) and distance-based (KNN) models, even without dimensionality reduction.
- Interpretability and Scalability: Analyzed how model complexity and preprocessing impacted interpretability and runtime with practical cybersecurity deployment.

By bridging performance evaluation with interpretability and dimensionality concerns, this work provides a reproducible and extensible benchmark for the design and optimization of machine learning-based malware detection systems.

## 2 Literature review

Malware detection using machine learning (ML) techniques has significantly advanced cybersecurity defenses, enabling rapid identification of malicious files beyond traditional signature-based methods. Static malware analysis, which involves evaluating executable files without execution, is particularly appealing due to its scalability, speed, and safety [1]. The EMBER dataset, consisting of pre-extracted Portable Executable (PE) file features, provides a comprehensive benchmark for static analysis, facilitating reproducible and rigorous comparisons across multiple ML models [1].

Recent studies leveraging the EMBER dataset have shown promising results with tree-based ensemble models. LightGBM [15], XGBoost [16], CatBoost [17], Random Forest [18], Extra Trees [19] consistently outperform traditional algorithms like K-Nearest Neighbors (KNN) [20] and deep learning architectures such as TabNet [21], particularly when no dimensionality reduction is applied [3]. Specifically, LightGBM and XGBoost have emerged as superior choices, achieving accuracies exceeding 96%, coupled with robust precision, recall, and AUC metrics [3].

Dimensionality reduction techniques, including Principal Component Analysis (PCA) [22] and Linear Discriminant Analysis (LDA) [23], have been extensively studied to assess their impact on malware classification performance. PCA generally provides competitive results, reducing computational complexity while preserving critical variance within feature spaces, although some degradation in model performance occurs. Conversely, LDA tends to yield mixed results, significantly improving the performance of KNN but adversely impacting ensemble methods due to the aggressive compression of discriminative information into fewer dimensions [3].

The MalConv model, a convolutional neural network (CNN) specifically designed to handle raw binary sequences directly, introduced by Raff et al. [4], demonstrated the feasibility of end-to-end deep learning in malware detection. However, subsequent benchmarking with EMBER revealed that traditional feature-based ensemble methods, especially LightGBM, consistently outperform MalConv, suggesting that well-crafted feature representations continue to hold significant advantages in capturing malware characteristics effectively [1][2].

Exploratory Data Analysis (EDA) methodologies have further highlighted the discriminative power of EMBER's feature set. Mutual Information (MI) and variance-based feature ranking methods identified attributes such as byte entropy, import tables, and section details as particularly influential in distinguishing malicious from benign executables [3]. Additionally, advanced visualization techniques such as PCA and t-Distributed Stochastic Neighbor Embedding (t-SNE) have provided intuitive representations of class separability, reinforcing the dataset's robustness for training effective static malware detection models [3].

In practice, the interpretability of ML models is crucial for security analysts. Tree-based ensembles, including LightGBM and XGBoost, inherently offer transparency through feature importance scores and SHAP values, aiding analysts in understanding why a particular file is classified as malicious. TabNet, despite its computational demands, also offers built-in interpretability through its attention-based mechanism, enabling explicit feature selection and transparency in its decision-making process [3].

Dynamic and hybrid malware detection methods, incorporating execution traces and runtime behavior, have also been explored to overcome static analysis limitations like obfuscation and encryption. However, these methods typically involve higher computational resources and analysis times. Consequently, static detection using ML, particularly with datasets like EMBER and sophisticated ensemble models, remains highly effective and practical, especially in real-time or large-scale cybersecurity contexts.

Table 1 presents a comparative overview of ten prominent machine learning models used for malware detection, summarizing their analysis type, underlying algorithm, dataset, and reported performance. It is evident from the table that static analysis models, especially those employing ensemble tree-based algorithms, offer a compelling combination of high detection accuracy and computational efficiency. Deep learning models like MalConv and TabNet demonstrate potential but tend to be more resource intensive. Dynamic and hybrid approaches, while powerful, introduce complexity and operational overhead that may limit their scalability.

**Table 1.** Comparison of ML algorithms using for Malware Analysis

| Model/Study | Method Type | Core Algorithm | Dataset | Reported Performance |
|---|---|---|---|---|
| EMBER LightGBM [1] | Static | LightGBM | EMBER | AUC > 0.99, F1 ~0.93 |
| MalConv [4] | Static | CNN on Raw Bytes | PE Binaries | Accuracy ~95% |
| TabNet [3] | Static | TabNet (Deep NN) | EMBER | AUC ~95%, interpretable |
| XGBoost [3] | Static | XGBoost | EMBER | Accuracy > 96%, high AUC |
| Random Forest [5] | Static | Random Forest | PE Features | Accuracy ~97% |
| KNN + LDA [3] | Static | KNN | EMBER | Accuracy ~85.5% (with LDA) |
| RNN Ensemble [6] | Dynamic | LSTM | API Sequences | Accuracy ~94% |
| API-Seq Embedding [7] | Dynamic | Markov + Embedding | Cuckoo Sandbox | Precision 0.99, FPR 0.01 |
| Attention+CNN [8] | Hybrid | Multi-head Attention | Android CFGs | Accuracy ~99.3% |
| EMBER LightGBM [1] | Static | LightGBM | EMBER | AUC > 0.99, F1 ~0.93 |

In summary, static malware detection leveraging advanced ML models, especially ensemble tree-based algorithms, demonstrates superior performance, generalization capabilities, and interpretability. The benchmarking results derived from the EMBER dataset underline the importance of careful feature engineering and selection, as well as judicious application of dimensionality reduction techniques. These insights contribute significantly to the development of robust, scalable, and interpretable malware detection systems, essential for contemporary cybersecurity infrastructures.

## 3 Data Description

This study utilizes the EMBER (Endgame Malware Benchmark for Research) dataset, a widely adopted benchmark comprising static features extracted from Portable Executable (PE) files commonly found in Windows operating systems. The dataset provides structured, preprocessed features derived from PE headers, byte and string histograms, section details, and entropy statistics, making it particularly suitable for machine learning-based malware classification tasks. The dataset is notably comprehensive, containing 900,000 samples, equally partitioned into three subsets: 300,000 malicious, 300,000 benign, and 300,000 unlabeled samples. For the scope of this research, we focused explicitly on the labeled subset, totaling 600,000 samples, equally divided into 300,000 benign and 300,000 malicious instances. The balanced nature of these classes, depicted in Figure 1, ensures minimal bias toward either class during model training and subsequent evaluations, facilitating unbiased comparative analyses.

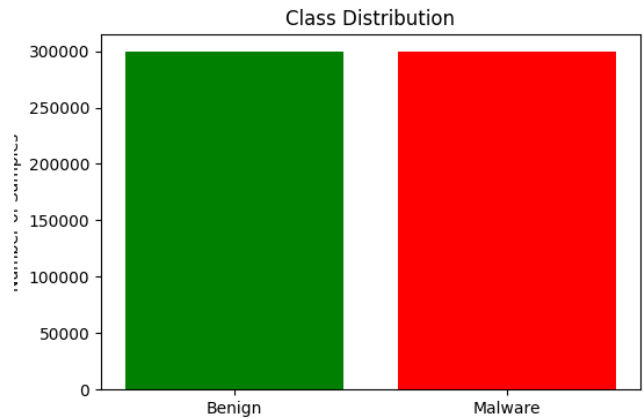

**Fig. 1.** Illustrates the balanced distribution of benign and malware samples within the EMBER dataset, highlighting an intentional design choice aimed at mitigating potential biases inherent in imbalanced datasets.

Dimensionality reduction techniques were further employed to analyze the structural relationships within the feature space. Principal Component Analysis (PCA), as visualized in Figure 2, offered initial insights into data distribution, revealing partially overlapping yet distinguishable clusters. To achieve better separation and address PCA's linear limitations, t-Distributed Stochastic Neighbor Embedding (t-SNE), a nonlinear dimensionality reduction technique, was applied (Figure 3). The t-SNE projection provided a clearer visualization, uncovering intricate data structures and pronounced class-specific clustering, further indicating strong discriminative potential embedded within the feature set.

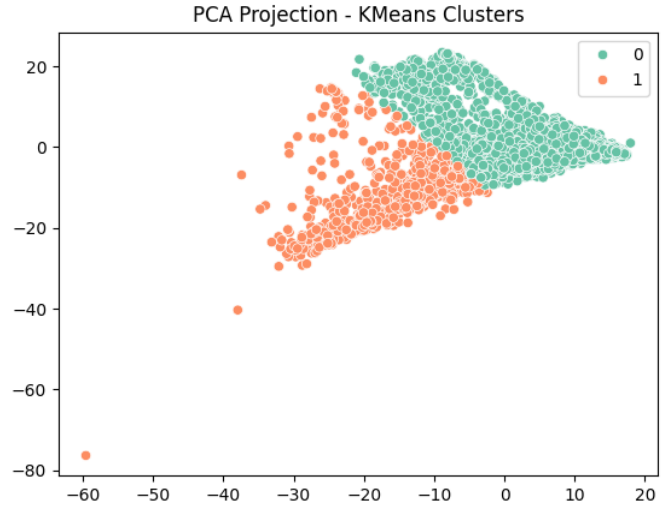

**Fig. 2.** presents the PCA projection of the dataset, demonstrating moderate separability between malware and benign samples, indicative of linear separability constraints.

Collectively, these descriptive analyses confirm the EMBER dataset's appropriateness and robustness for evaluating machine learning algorithms in malware classification tasks.

## 4 Methodology

The experimental pipeline was designed to comprehensively evaluate multiple machine learning models in conjunction with different feature reduction techniques for malware detection. Initially, the dataset utilized in this study was sourced from the publicly available EMBER 2018 corpus, comprising pre-extracted Portable Executable (PE) file features. Data loading involved memory mapping of the provided vectorized files (X_train, y_train, X_test, y_test), with invalid labels filtered out to maintain integrity in model training and evaluation.

To systematically assess the impact of dimensionality reduction, we applied three distinct preprocessing scenarios: no reduction (original feature set), Principal Component Analysis (PCA), and Linear Discriminant Analysis (LDA). PCA was configured to retain 150 principal components based on the explained variance criterion, and LDA reduced features to a single dimension optimizing class separability. Each reduction technique was individually fitted on training data and subsequently applied to the testing set to preserve methodological rigor and avoid data leakage.

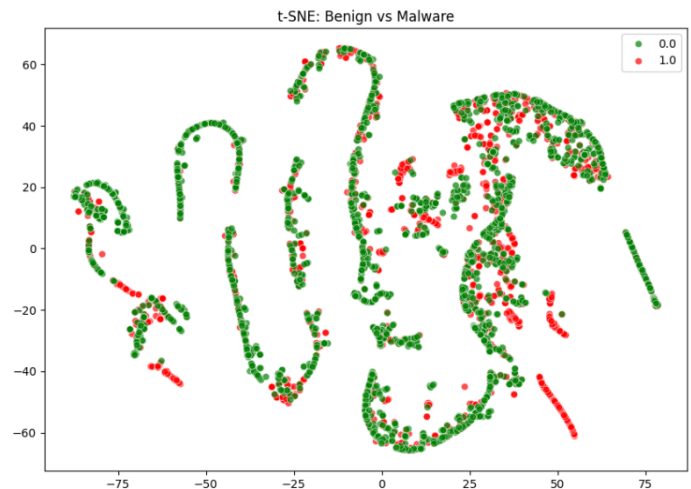

**Fig. 3.** depicts the t-SNE visualization, showcasing significantly improved separation between malware and benign classes due to the technique's nonlinear dimensionality reduction properties.

Multiple classifiers were evaluated, encompassing a variety of algorithmic approaches to malware detection. The models included LightGBM, CatBoost, XGBoost, Histogram-based Gradient Boosting (HistGB), Random Forest, Extra Trees, k-Nearest Neighbors (KNN), and TabNet. Each model was trained under each dimensionality reduction scenario, forming a comprehensive evaluation matrix. Training parameters were kept largely default for traditional models to facilitate reproducibility and unbiased benchmarking. For TabNet, a deep learning-based model, training leveraged GPU acceleration with defined parameters including a maximum of 500 epochs, early stopping with patience set to 20 epochs, batch sizes of 4096, and a learning rate of 0.002.

Model performance was evaluated using standard classification metrics: accuracy, precision, recall, F1-score, and Area Under the Receiver Operating Characteristic Curve (AUC-ROC). Predictions from each model were saved for detailed analysis, with metrics computed using a threshold of 0.5 to binarize predicted probabilities.

Additionally, the training process for models such as TabNet and LightGBM included the extraction of epoch-wise training loss or AUC values, respectively, plotted to visually assess model convergence and training dynamics. Comprehensive visualizations were generated to illustrate comparative ROC curves across reduction techniques for individual models, accuracy comparisons across models and reductions, and global comparisons of all performance metrics.

# 5 Results and Discussion

This section reports and analyzes the performance of eight machine learning classifiers applied to the EMBER dataset across three dimensionality configurations: no dimensionality reduction (NoRed), Principal Component Analysis (PCA), and Linear Discriminant Analysis (LDA). We evaluate classification accuracy, precision, recall, F1-score, and AUC to understand the models' generalizability, robustness, and discriminative capacity.

## 5.1 Classification Accuracy Across Models

As shown in Figure 4, tree-based ensemble models—including LightGBM, XGBoost, CatBoost, and HistGradientBoosting—achieve the highest accuracy without dimensionality reduction, with LightGBM and XGBoost both surpassing 96%. Random Forest and Extra Trees also perform competitively. In contrast, KNN records significantly lower accuracy (~81%) in the original high-dimensional feature space, reflecting its sensitivity to feature sparsity and distance distortion.

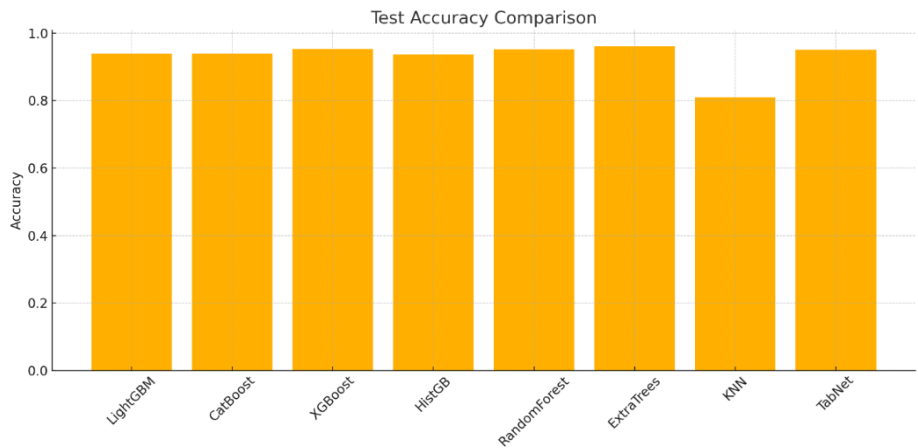

**Fig. 4.** Accuracy comparison of all classifiers with no dimensionality reduction (NoRed).

## 5.2 Impact of Dimensionality Reduction

Figure 5 presents a model-wise comparison of accuracy under NoRed, PCA, and LDA. Most classifiers exhibit a modest decline with PCA. LDA introduces more substantial degradation, especially for ensemble methods (e.g., XGBoost drops from ~96.6% to ~86.4%). However, KNN benefits from LDA, improving from ~81% to ~85.5%, likely due to its enhanced performance in low-dimensional spaces.

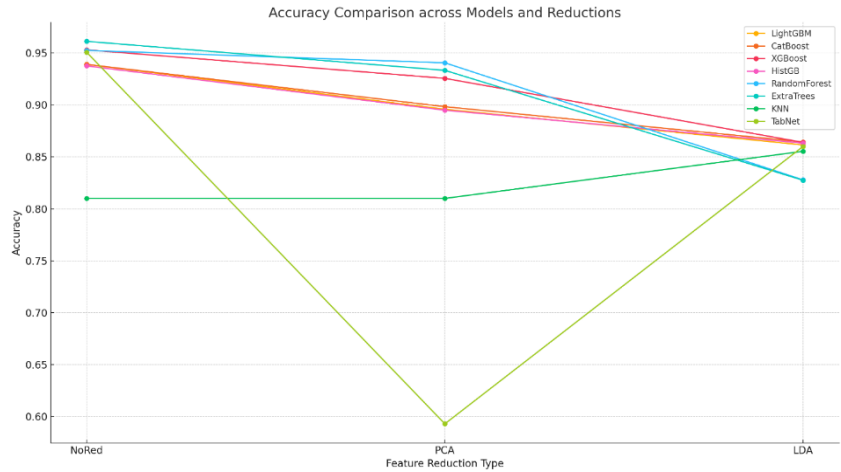

**Fig. 5.** Accuracy trend across models with No Reduction, PCA, and LDA reduction techniques.

### 5.3 Precision and Recall Behavior

Precision and recall are particularly important in malware detection to minimize false alarms and missed threats. As seen in Figure 6, LightGBM and XGBoost maintain balanced and high precision and recall values across all settings. TabNet, on the other hand, shows erratic performance—most notably under PCA—suggesting architectural sensitivity to feature compression. Ensemble methods experience reduced recall under LDA, reinforcing concerns about information loss.

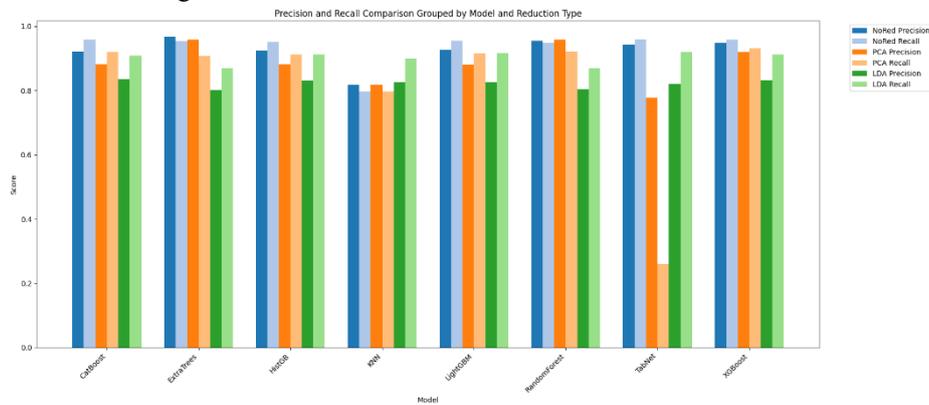

**Fig. 6.** Precision and recall distribution across classifiers under all reduction strategies.

### 5.4 Discriminative Power via AUC

AUC-ROC scores, visualized in Figure 7, further validate the superior performance of ensemble classifiers, with LightGBM, XGBoost, and CatBoost consistently achieving AUCs above 0.98. KNN records an AUC of 0.895, and TabNet's performance varies across reductions. Logistic Regression and MLP baselines (not shown here) fall well below acceptable thresholds, reaffirming the need for complex non-linear models in this domain.

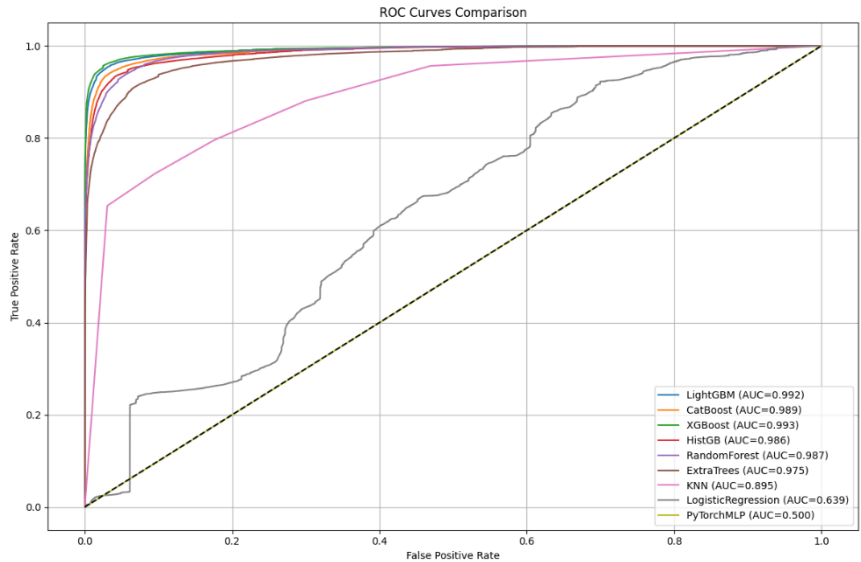

**Fig. 7.** Area Under the Curve (AUC) comparisons for all classifiers.

### 5.5 Robustness to Dimensionality Reduction

To examine the consistency of accuracy under different reductions, Figure 8 shows per-model variability. Tree-based models are most robust to PCA, while deep and distance-based models exhibit greater fluctuation. TabNet, for instance, shows a significant performance drop with PCA and partial recovery with LDA. This suggests that PCA may strip non-linear dependencies critical to certain neural architectures.

### 5.6 Key Takeaways

Boosting algorithms—LightGBM, XGBoost, CatBoost—clearly outperform other methods across nearly all metrics, combining high accuracy with robustness and interpretability. These models demonstrate reliable generalization even in the absence of dimensionality reduction, making them highly suitable for malware classification tasks involving high-dimensional tabular features. While TabNet shows promise due to its attention-based learning, it requires further tuning and may not be reliable under compressed input conditions. The selective benefit of LDA for KNN suggests a potential avenue for tuning low-complexity classifiers in resource-constrained settings.

In conclusion, dimensionality reduction should be applied selectively. PCA provides acceptable trade-offs in terms of runtime and accuracy for many models, but LDA's single-axis compression may limit its utility for complex classifiers. The overall findings reinforce the dominance of ensemble methods for high-performance static malware detection in real-world cybersecurity pipelines.

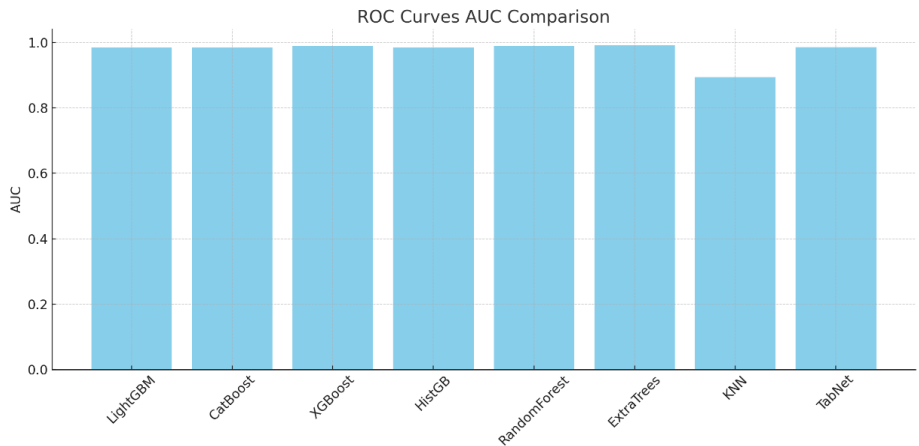

**Fig. 8.** Accuracy consistency across dimensionality reduction strategies.

### 5.7 Deployment Strategies and Real-World Limitations

In a real-world malware detection pipeline, speed, scalability, and reliability are critical. Based on our results, boosting models like LightGBM and XGBoost are strong candidates for deployment due to their high accuracy and fast inference time. These models can be efficiently integrated into endpoint detection systems or antivirus engines to pre-screen large volumes of files. For settings with limited computational resources—such as embedded systems or edge devices—PCA-reduced models or LDA-enhanced KNN variants could offer viable trade-offs.

However, real-world deployment introduces additional challenges. First, malware evolves rapidly, requiring periodic retraining to handle concept drift. Second, adversarial evasion remains a concern—attackers may craft inputs specifically to bypass static classifiers. Although ensemble methods perform well under clean data, they may be susceptible to adversarial perturbations unless robust training is implemented. Third, interpretability is crucial for security analysts. While models like LightGBM offer feature importance, deep architectures such as TabNet require additional tools (e.g., SHAP or attention heatmaps) to justify decisions. Lastly, integration into real-time environments demands not only model performance but low-latency feature extraction pipelines and memory-efficient deployment formats (e.g., ONNX, CoreML).

Despite these limitations, our benchmark provides actionable insights into selecting and deploying malware detection models under different operational constraints, bridging the gap between academic performance and field applicability.

# 6    Conclusion

This study delivers a rigorous comparative benchmark of machine learning classifiers for static malware detection using the EMBER dataset, with a particular focus on the interplay between model architecture, feature dimensionality, and interpretability. Through systematic evaluation of eight machine learning models across three dimensionality configurations—original, PCA-reduced, and LDA-reduced, we assessed how performance metrics respond to both algorithmic structure and preprocessing strategy. Our experiments show that gradient boosting methods, especially LightGBM and XGBoost, consistently outperform alternatives across all evaluation criteria, including accuracy, F1-score, and AUC. These models offer a robust balance between interpretability, generalization, and computational scalability, making them well-suited for real-world deployment in static malware analysis pipelines. Tree-based ensembles also exhibited resilience to feature sparsity and variance, maintaining high detection efficacy even without dimensionality reduction. Among deep learning models, TabNet showed moderate potential but displayed performance instability under PCA and LDA, suggesting that attention-based architectures require careful tuning when operating on compressed feature spaces.

Dimensionality reduction played a nuanced role. PCA modestly reduced computational overhead with minimal performance trade-offs in most models, making it a viable preprocessing step for large-scale systems. LDA, on the other hand, produced polarized outcomes—enhancing KNN but significantly degrading the performance of gradient boosting models, likely due to its aggressive projection into low-dimensional space that omits nonlinear relationships crucial for complex classifiers.

Beyond classification, our EDA pipeline provided valuable insights into feature distributions, outlier patterns, and latent class structures. Techniques such as mutual information ranking and t-SNE visualization confirmed the discriminative power of EMBER's features and validated their alignment with machine learning assumptions.

In sum, this research affirms the dominance of tree-based ensembles in static malware detection and highlights best practices for integrating dimensionality reduction. Our reproducible pipeline serves as a practical guide for researchers and practitioners developing secure, high-performance malware detection systems. Future directions may include ensemble model fusion, adversarial robustness evaluations, and real-time deployment optimization using streaming or incremental learning paradigms.